\documentclass[superscriptaddress,twocolumn,showpacs,prl]{revtex4-1}
\usepackage{graphicx}
\usepackage{amsmath}
\usepackage{amssymb}

\begin{document}
\title{Isolated quantum heat engine}
\author{O. Fialko and D. W. Hallwood}
\affiliation{Centre for Theoretical Chemistry and Physics and NZIAS, Massey University, Private Bag 102904, North Shore, Auckland 0745, New Zealand}

\begin{abstract}
We present a theoretical and numerical analysis of a quantum system
that is capable of functioning as a heat engine. This system could be
realized experimentally using cold bosonic atoms confined to a double
well potential that is created by splitting a harmonic trap with a
focused laser. The system shows thermalization, and can model a
reversible heat engine cycle. This is the first demonstration of the
operation of a heat engine with a finite quantum heat bath.
\end{abstract}

\maketitle

Recent developments in modeling the quantum dynamics of isolated
systems \cite{peres84}  have provided a quantum understanding of thermalization.
However, they rely on experimentally challenging systems. This has
meant that development of a quantum understanding of heat engines has
not been possible. In this Letter, we consider a model of cold bosonic
atoms confined to a double well potential. Our analysis shows that this system exhibits thermalization when
one well is initially more energetic than the other, and furthermore shows
that the system can perform a heat engine cycle. The system could be
realized experimentally with present technology.

The state of a quantum system evolves via the application of a unitary operator. Consequently, all processes are reversible and until recently 
it was not clear how a quantum system could reach thermal equilibrium. Srednicki suggested a solution called the eigenstate thermalization hypothesis 
(ETH)~\cite{srednicki94}, in which all eigenstates have the properties of a thermal state.
Recently, this hypothesis was tested against other hypotheses by Rigol \emph{et al.} \cite{rigol08} using hard core bosons confined to a lattice and found to be the only one that agreed. 

Several experimentally simpler schemes have been studied. Ponomarev \emph{et al.} demonstrated thermal equilibration when two isolated systems with different temperatures were coupled through interactions between atoms in the different systems~\cite{ponomarev11}. Chianca \emph{et al.} allowed tunneling of atoms between the systems in a four site lattice configuration~\cite{olsen11}. However no simple measure of temperature could be obtained meaning it could not be developed into a heat engine. 
A quantum description of a heat engine has been produced in Refs.~\cite{kosloff00} where the system is coupled to a classical heat bath.

In this Letter, a quantum analysis of a heat engine is performed where both the system and heat bath are modeled quantum mechanically and atoms can 
tunnel between wells. Each well is described by two energy levels allowing temperature to be defined. One  well acts as the system and the other 
serves as the heat bath. Heat is added and removed by tunneling of atoms between the wells. The volume of the wells are changed by modifying the 
harmonic confinement and the barrier position. All processes are done at a rate that provides good reversibility of the engine. 
The system is modeled using both the truncated Wigner approach (TWA) to calculate the system for thousands of atoms and the accuracy is checked using 
calculations of the full quantum dynamics (FQD) for tens of atoms. Both results agree well showing an increase in number reduces fluctuations.

\begin{figure}
\begin{center}
\includegraphics[width=7cm]{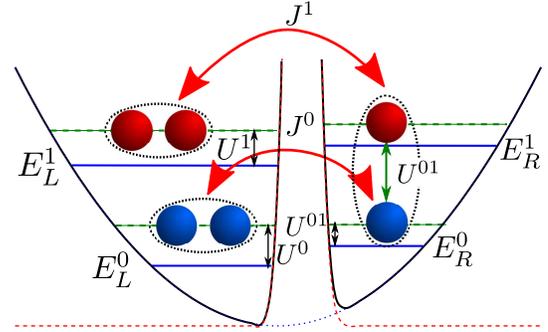}
\end{center}
\caption{Schematic of a double well created by splitting a harmonic potential with a focused laser. The diagram shows the possible tunneling and how the energy levels change due to the interactions.}
\label{fig:system}
\end{figure}

$N$ bosons confined to a double well trapping potential, $V_{dw}(x)$, are described by the Hamiltonian
\begin{eqnarray}
\nonumber
\hat{H}&=& \int dx \left[ \frac{-\hbar^2}{2m} \nabla \hat{\psi}^{\dag}(x) . \nabla \hat{\psi}(x) + V_{dw}(x) \right] \\
&+& g\int dx \int dx' \hat{\psi}^{\dag}(x)\hat{\psi}^{\dag}(x') \delta (x-x')\hat{\psi}(x')\hat{\psi}(x).
\label{eq:2welsecquant}
\end{eqnarray}
At low temperatures only the lowest laying single particle states are populated. Here we take into account the first two states as depicted in Fig.~\ref{fig:system}.  Therefore the field operators can be described in terms of the four localized single particle functions,
$\hat{\psi}(x)= \sum_{l=0}^1 \left( \phi_L^{l}(x) \hat{b}_L^{l} + \phi_R^{l}(x) \hat{b}_R^{l} \right)$,
where $\hat{b}_r^l$ are the bosonic annihilation operators of an atom in well $r$ and energy level $l$ and described 
by the single particle functions $\phi_r^l$. This leads to the two-band Hubbard Hamiltonian \cite{carr07, carr10}
\begin{eqnarray}
\nonumber
\hat{H}&=&-\sum_{ r\ne r' ,l} J^l \hat{b}_r^{l\dagger}\hat{b}_{r'}^{l}+\sum_{r,l}U^l\hat{n}_r^l (\hat{n}_r^l-1)
+\sum_{r,l}E_r^l \hat{n}_r^l \\
&&+U^{01}\sum_{r,j\ne l'}(2\hat{n}_r^l \hat{n}_r^{l'}+\hat{b}_r^{l\dagger}\hat{b}_{r}^{l\dagger}
\hat{b}_r^{l'}\hat{b}_{r}^{l'}),
\label{eq:2wel2levelBHM}
\end{eqnarray}
where we have ignored interactions between atoms in different wells. The ground and first excited state energies are $E_r^l = \int dx \phi_r^{l*}(x) \left(- \frac{\hbar^2}{2m}\nabla^2+ V_{dw}(x) \right) \phi_r^{l}(x)$. The tunnel coupling between the wells is $J^l = \int dx \phi_L^{l*}(x) \left(- \frac{\hbar^2}{2m}\nabla^2+ V_{dw}(x) \right) \phi_R^{l}(x)$. The interaction between atoms in the same well and on the same energy level is $U^l = g \int dx |\phi_r^{l}(x)|^4$, and on different energy levels is $U^{01} = g \int dx |\phi_r^{0}(x)|^2 |\phi_r^{1}(x)|^2$. This term also leads to atoms changing energy levels. 

We consider a harmonic potential with oscillator frequency $\omega_0$, which is split by a focused laser beam located at $x_0$ from the center of 
the harmonic potential and described by a Gaussian potential $V_0 \exp(-(x-x_0)^2/2\sigma^2)$. The barrier height $V_0 = 10\hbar\omega_0$, 
with width $\sigma=0.1 l_{ho}$, where $l_{ho}= \sqrt{\hbar/m\omega_0}$ is the harmonic oscillator frequency. For a symmetric well, localized functions 
representing the energy levels in the different wells were calculated from the single particle eigenstates of the system. 
This gives $J^0/\hbar\omega_0 = 0.153$, $J^1/\hbar\omega_0 = 0.226$, $E_r^0/\hbar\omega_0 = 1.37$ and $E_r^1/\hbar\omega_ 0= 3.31$. 
The interaction terms can be calculated from the integrals above and the interaction coupling, $g$. 
The interaction coupling can be varied by the Feshbach resonance \cite{bloch08} and for our purpose we use $U^0/\hbar\omega_0 = 2/N$, 
$U^1=3U^0/4$ and $U^{01} = U^0/2$. During the heat engine cycle only $E_r^l$ are modified, because other terms are approximately constant.


\begin{figure}[t]
\begin{center}
\includegraphics[width=7cm]{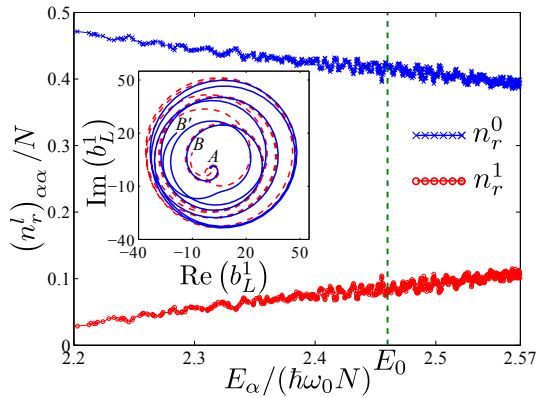}
\end{center}
\caption{The expectation values of the occupation numbers in an energy window near the mean energy of the system $E_0$ for $N=40$. 
The intersection with the mean energy of the system (dotted line) gives approximate values of the thermalized values. This supports
the ETH.
Inset: The dynamics of the field $b_L^1$ from $t=0$ to $t=10\omega_0^{-1}$ in the symmetric double well for two slightly different initial conditions. 
The two trajectories starting at $A$ and ending at B(B') are not periodic, they gradually diverge and
cover finite space. This demonstrates chaos in our system \cite{katok03}.}
\label{Fig.chaos}
\end{figure}

Thermalization of the system can be understood by first writing the initial wavefunction in terms of the eigenstates of the Hamiltonian,
$|\Psi(0) \rangle = \sum_{\alpha} C_{\alpha} |\phi_{\alpha}\rangle$,
where ${|\phi_{\alpha}\rangle}$ are the eigenstates with energies ${E_{\alpha}}$ and $C_{\alpha} = \langle \phi_{\alpha} | \Psi(0)\rangle$. The time evolution is given by,
$|\Psi(t) \rangle = e^{-i H t/\hbar} |\Psi(0) \rangle$.
The expectation value of an observable $\cal{O}$ is thus given by
\begin{equation}
\langle \hat{\cal{O}} \rangle = \sum_{\alpha\beta} C_{\alpha}^* C_{\beta} e^{i(E_{\alpha}-E_{\beta})t/\hbar} \cal{O}_{\alpha\beta},
\label{eq:A_exp}
\end{equation}
where $\cal{O}_{\alpha\beta}=\langle \phi_{\alpha} |\hat{\cal{O}}| \phi_{\beta} \rangle$. As stated above, ETH describes the long time dynamics of 
the system~\cite{srednicki94}. The eigenstates of the system that make up the initial state have the properties of thermal state. 
After a considerable time, the off-diagonal terms in Eq.~(\ref{eq:A_exp}) average to zero. 
This is called the long time average (LTA), which is a weighted average of the eigenstates and is independent of time,
$\overline{\langle \hat{\cal{O}} \rangle} = \sum_{\alpha} |C_{\alpha}|^2 \cal{O}_{\alpha\alpha}$. 
If the system is thermalized, then the long time average should be equal to an ensemble average, which is given 
by a microcanonical average over a narrow energy interval around the mean energy of the system,
$\langle \hat{\cal{O}} \rangle_{E_0}=1/{\cal{N}}\sum_{\alpha, E_0\pm\Delta E/2}\cal{O}_{\alpha\alpha}$.
Here $\cal{N}$ is the number of states in the energy interval $\Delta E$, $E_0=\sum_{\alpha}|C_{\alpha}|^2 E_{\alpha}$
is the mean energy of the system. We investigate the number of atoms in each of the energy levels. Therefore, the following equality 
is expected to be valid for thermalization  
\begin{equation}
\lim_{t\rightarrow\infty}\langle \hat{n}_r^l(t) \rangle = \overline{\langle \hat{n}_r^l \rangle} = \langle \hat{n}_r^l \rangle_{E_0}.
\label{Eq:equal}
\end{equation}
According to the ETH knowing a single eigenstate is sufficient to compute averages,
${\cal{O}}_{\alpha\alpha}\approx \langle \hat{\cal{O}} \rangle_{E_\alpha}$ \cite{srednicki94,rigol08}. 
For $\hat{\cal{O}}=\hat{n}_r^l$ this is shown in  Fig.~\ref{Fig.chaos}. The ETH also assumes initial conditions narrow in energy.
For the initial state we use (see below) 
$(\langle \hat{H}^2 \rangle-\langle \hat{H} \rangle^2)^{1/2}/\langle \hat{H} \rangle\sim 0.1$.


\begin{figure}[t]
\begin{center}
\includegraphics[width=8cm]{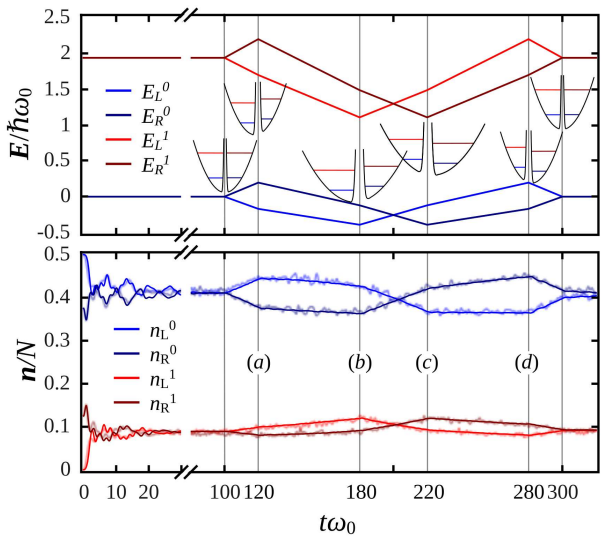}
\end{center}
\caption{(Color online) The heat engine cycle. TOP: the energy of the localized states in the double well system as the system evolves. BOTTOM: the population expectation value of each of the energy levels for TWA (dark lines) and FQD (faint lines). The initial state is $|N/2,3N/8,0,N/8 \rangle$ and is first allowed to thermalize. The tunneling and interaction strengths are kept constant throughout the simulations at  $J^0/\hbar\omega_0 = 0.153$, $J^1/\hbar\omega_0 = 0.226$,  $U^0/\hbar\omega_0 = 2/N$, $U^1=3U^0/4$ and $U^{01} = U^0/2$.}
\label{Fig.heat_engine}
\end{figure}


There are no general arguments supporting the ETH. There are proofs that quantum systems, whose classical
analogues are chaotic, satisfy the ETH in the semiclassical limit \cite{voros79}.
It is not computationally tractable to calculate the classical counterpart of our system for hundreds of atoms,
therefore we look at the mean-field version of our system first. In this approach we replace the operators with 
complex numbers $\hat{b}_r^l\rightarrow b_r^l$ in the Heisenberg equations of motion
$\frac{d \hat{b}_r^l}{d t}=\frac{i}{\hbar}[\hat{H}, \hat{b}_r^l]$ and get
\begin{eqnarray}
\nonumber
i\frac{d b_{\rm r}^l}{d t}=-J^lb_{\rm \bar{r}}^l+2U^l |b_{\rm r}^l|^2 b_{\rm r}^l+(E_r^l-U^l)b_{\rm r}^l \\
+ 2U^{01}|b_{\rm r}^{1-l}|^2 b_{\rm r}^l +2U^{01}b^{l\ast}_{\rm r} b_{\rm r}^{1-l} b_{\rm r}^{1-l},
\label{Eq.Heisenberg}
\end{eqnarray}
where $l=\{1,2\}$, $r=\{L,R\}$ and $\bar{L}=R$,  $\bar{R}=L$. The dynamics of the system is chaotic (see e.g. \cite{chaos01}).
For example the imaginary and real parts of $b_L^1$ shown in the inset of Fig. \ref{Fig.chaos} 
clearly exhibit chaotic behavior. It follows that the trajectories are sensitive to initial conditions. 
We apply the truncated Wigner approximation, which is a powerful tool for the investigation of quantum dynamics in interacting many body 
systems \cite{blakie08}. This amounts to solving 
Eq.~(\ref{Eq.Heisenberg}) with different initial conditions, which accurately samples quantum noise in the system~\cite{ashton09}, 
and averaging afterwards. Since the trajectories of the system are chaotic,
this might lead to equilibration within the TWA. For a coherent state, we sample the initial conditions as $b=b_0 + \frac{1}{2}(\nu_1 + i\nu_2)$, where $|b_0|^2=n+1/2$ and $\nu_i$ are Gaussian random variables. The coherent state corresponds to a condensed state of bosons. Another possibility is a Fock state, which is sampled as $b=(p+q\nu)e^{i2\pi\xi}$, where $\nu$ is a Gaussian random variable and $\xi$ is a uniform random variable in the interval $[0,1]$, $p=\frac{1}{2} \left(2N+1+2\sqrt{N^2+N}\right)^{1/2}$ and $q=\frac{1}{4p}$.
The initial configuration of  $N=10^4$ atoms used is $n^0_{\rm L}=N/2$, $n^0_{\rm R}=3N/8$, $n^1_{\rm L}=0$, $n^1_{\rm R}=N/8$, so the right well is 
more energetic than the left well. The number of atoms in the excited state is small so the temperature is low enough to satisfy the two band approximation 
(see later). The system undergoes thermalization up to a time $t = 100\omega_0^{-1}$ (see Fig. \ref{Fig.heat_engine}). 
The expectation values of the occupation number of particles approach a constant value, which is independent of the initial state (being coherent state or Fock state) and agrees well with ETH.
We find the system has thermalized with values $n_r^1/N \approx 0.089$ for both the left and right wells, and $n_r^0+n_r^1=N/2$.

Another method simulates the FQD by creating the Hamiltonian in the Fock basis of localized functions and propagates 
using the unitary matrix $U(t) = \exp(-iHt/\hbar)$. To make this approach computationally tractable the system is limited 
to 40 atoms. However, this is enough to verify the results of the TWA. Figure \ref{Fig.heat_engine} shows that the FQD 
gives values that oscillate around the TWA results. These oscillations reduce for larger numbers of atoms and for the 
$N=10^4$ atoms we expect good agreement with the TWA. 
{ Eq.~(\ref{Eq:equal}) is verified and
compared with $n_r^1$}, which we average between times $t \omega_0 = 80$ to $t \omega_0 = 100$ to account for 
the oscillations, and gives good agreement, $n_r^1/N \approx \overline{\langle \hat{n}_r^1 \rangle}/N \approx \langle \hat{n}^1_r \rangle_{E_0}/N \approx 
0.089$.

\begin{figure}
\begin{center}
\includegraphics[width=7cm]{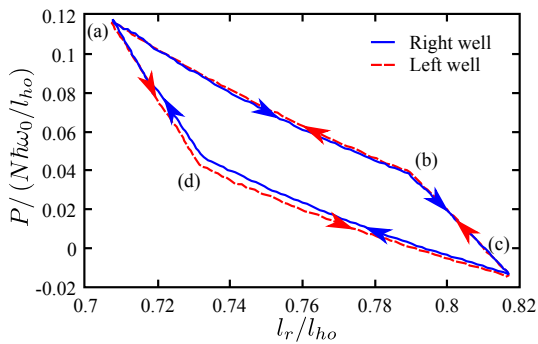}
\end{center}
\caption{Pressure-volume plot for the right and left wells. Each well produces work proportional to the area enclosed by the curves.}
\label{fig:PV}
\end{figure}

The heat engine scheme is similar to the Otto Cycle \cite{otto}, and was chosen to be experimentally simple. 
Heat is transferred from one well to another via tunneling of atoms. The work is done on the trapping potential by the expansion of atoms 
in a well when its size is slowly increased.
Here the right well is the engine while the left well serves as the heat bath, and this corresponds to 
the clockwise solid line in Fig.~\ref{fig:PV}. The heat engine process starts at $t=100\omega_0^{-1}$.
$(a) \rightarrow (b)$: the laser barrier starts at $x_0=0.2/l_{ho}$ and the harmonic confinement is adiabatically reduced to $\omega=0.8\omega_0$, 
so the populations of the localized functions remain roughly constant. 
This relates to the power stroke in the Otto cycle and work is done by the engine. 
$(b) \rightarrow (c)$: the barrier moves to the left, $x_0=-0.2/l_{ho}$, leading to a rapid change 
in the populations. This is the heat rejection stage of the Otto cycle and heat is extracted from the right well.  
$(c) \rightarrow (d)$: the harmonic confinement is adiabatically increased and the populations of the localized functions again remain roughly constant. 
This is the compression stage in the Otto cycle and work is done on the engine. 
$(d) \rightarrow (a)$: the final stage sees the barrier move to the right and relates to the combustion stage of the Otto cycle. 
Here heat is added to the engine. While the volume does change in the steps $(b) \rightarrow (c)$ and $(d) \rightarrow (a)$, 
it is small relative to the adiabatic processes. 
The slight asymmetry of the evolution is due to the finite evolution time of the system, which was necessary to limit numerical inaccuracy.

If a system is thermalized, it reaches a thermal state, which is described by the thermal density matrix $\rho_T =\exp{(-\hat{H}/k_B T)}/ 
{\rm Tr}\{\exp{(-\hat{H}/k_B T)}\}$ \cite{gardiner04}.
The temperature $T$ is chosen to best fit the data, and is used to calculate $ n^l_r = \mbox{Tr}
(\rho_T \hat{n}_r^l)$. At $t=100\omega_0^{-1}$ the system is thermalized with $T_1\approx 2.6 \hbar\omega_0/k_B$.
We notice here that the two wells are not always in thermal equilibrium during the engine cycle,
since we require the atoms flow from one well into another. However, the system reaches thermal equilibrium at $t=200\omega_0^{-1}$
with $T_2\approx 2.2 \hbar\omega_0/k_B$ and completes the cycle with $T\approx T_1$.
The thermal energy, $k_BT$, is of the same order as the gap between 
the ground and first excited single particle states, $E^1_r-E^0_r\approx 2\hbar\omega_0$, which validates the two-level approximation.

The pressure and volume (or length for a quasi-one dimensional system) in each well can be used to demonstrate the efficiency of the engine. 
If we assume the wells can be approximated by harmonic oscillators, then we can use the harmonic oscillator length as the length scale of the system, 
so $l_r=\sqrt{\hbar/m\omega_r}=\hbar/\sqrt{m(E_r^1-E_r^0)}$.  
For simplicity, we use the pressure for an ideal gas, $P=2E_r/3V$  ~\cite{pitaevskii03},   
where $E_r = n_r^0 E_r^0 + n_r^1 E_r^1$ is the kinetic energy of a well, 
and taking $V\approx 2l_r$, we get $P = E_r/3l_r$. 
The pressure and volume of the left and right wells are plotted in Fig.~\ref{fig:PV} and we see the hysteresis of the heat engine, which shows work has been done.
The whole isolated system cannot produce work, 
since the two processes in Fig.~\ref{fig:PV} cancel each other. This is consistent with the second law of thermodynamics.

The entropy of a well is another quantity along with temperature, which can be measured to verify that the system has come into thermal equlibrium. 
It is given by ${\cal{S}} = - k_B\mbox{Tr} \left\{ \rho_r \log\rho_r \right\}$, 
where $\rho_r$ is the reduced density matrix traced over the other well. As expected, it starts from zero at $t=0$ and
approaches ${\cal{S}}\approx 0.1 k_B N$ at $t=100\omega_0^{-1}$.
The entropy and other thermodynamic quantities can be measured from scanning the density profile of a quantum gas 
\cite{ho10}. A focused electron beam with extremely high resolution of 150 nm can be used  for this purpose 
\cite{ott08}.

The efficiency can be calculated as $\eta={\cal W}/{\cal Q}_{\rm in}$, where work done is ${\cal W}\approx\sum_{l=0,1}\int n_R^l d E_R^l$ 
and heat added to the engine is ${\cal Q}_{\rm in}\approx\sum_{l=0,1}\int \epsilon_R^l d n_R^l$, 
where $\varepsilon_r^l = E_r^l + 2U^l n_r^l + 2U^{01} n_{r}^{1-l} $ 
is the energy of a single particle shifted by the mean-field contribution from other particles in the same level \cite{kosloff00}. 
We obtain $\eta\approx 0.14$. This is close to the Carnot efficiency $\eta\approx 1-T_2/T_1\approx 0.15$.

To experimentally realize the heat engine cycle, a magnetic trap with radial and axial trapping frequencies 
of $\omega_{\perp}/2\pi=4$kHz and $\omega_{0}/2\pi=40$Hz can be used to confine atoms in a 1D trapping potential. 
In this case the ratio of the corresponding  oscillator lengths is $l_{ho}/l_{\perp}\approx 10$,
such that the width of the focused laser beam is $\sigma\approx l_{\perp}$.
For $^{7}$Li atoms this gives  $\sigma\approx 0.6\mu m$. A narrow  laser beam was recently reported with $\sigma\approx 0.7 \mu m$  
and positioning the beam  to a lateral precision of $0.05 \mu m$ \cite{bloch11}. 
Taking the 1D interaction strength $g=2\hbar^2 a_s/ma_{\perp}^2$ \cite{pitaevskii03} yields the estimate 
for the two-body scattering length $a_s\sim 100$nm$/N$. The scattering length for $^7$Li atoms was achieved 
as small as $\sim 10^{-4}$nm \cite{hulet09}, therefore the number of particles should not exceed $N \sim 10^6$.   
For such a realistic setup the full cycle of the engine  will take $\sim 1 s$. 

In conclusion, we have demonstrated thermalization of cold bosonic atoms confined to a double well potential,
which is described by the two-band Hubbard model. We have shown that this can be utilized for realization of an isolated finite quantum engine, 
where one well produces work and another well serves as the heat bath. If realized experimentally, this will demonstrate the first isolated 
quantum engine with finite heat bath. 

We thank Murray Olsen and Ho Tsang Ng for useful discussions. This work was supported by the
Marsden Fund (contract No. MAU0910 and MAU0706), administered by the Royal Society of New Zealand.\\



\end{document}